\documentclass[runningheads]{llncs}
\usepackage[table,xcdraw]{xcolor}
\usepackage{graphicx}
\usepackage{graphicx}
\usepackage{xcolor}
\usepackage{float}

\begin{document}
\title{Influence of Autoencoder Latent Space on Classifying IoT CoAP Attacks}
%
%
\author{Mar\'ia Teresa Garc\'ia-Ord\'as\inst{1}\and
Jose Aveleira-Mata\inst{2} \and
Isa\'ias Garc\'ia-Rodr\'igez\inst{1} \and
Jos\'e Luis Casteleiro-Roca\inst{3}\and
Mart\'in Bay\'on-Gutierrez\inst{1}\and
H\'ector Alaiz-Moret\'on\inst{2}\
}
\authorrunning{ Garc\'ia-Ord\'as et al.}

\institute{
University of Le\'on, Department of Electrical and Systems Engineering,
\\ Escuela de Ingenier\'ias, Campus de Vegazana, 24071 Le\'on, Spain\\
\email{\{mgaro,igarr,martin.bayon\}@unileon.es}
\and
Research Institute of Applied Sciences in Cybersecurity (RIASC)  MIC. University of Le\'on, 24071 León.\\
\email{\{jose.aveleira,hector.moreton\}@unileon.es}
\and
University of A Coru\~na, CTC, Department of Industrial Engineering, CITIC\\ Avda. 19 de febrero s/n, 15405, Ferrol, A Coru\~na, Spain\\
\email{\{jose.luis.casteleiro\}@udc.es}}

\maketitle              
\begin{abstract}

The Internet of Things (IoT) presents a unique cybersecurity challenge due to its vast network of interconnected, resource-constrained devices. These vulnerabilities not only threaten data integrity but also the overall functionality of IoT systems. This study addresses these challenges by exploring efficient data reduction techniques within a model-based intrusion detection system (IDS) for IoT environments. Specifically, the study explores the efficacy of an autoencoder’s latent space combined with three different classification techniques. Utilizing a validated IoT dataset, particularly focusing on the Constrained Application Protocol (CoAP), the study seeks to develop a robust model capable of identifying security breaches targeting this protocol. The research culminates in a comprehensive evaluation, presenting encouraging results that demonstrate the effectiveness of the proposed methodologies in strengthening IoT cybersecurity with more than a 99\% of precision using only 2 learned features.
\end{abstract}

\section{Introduction}

The Internet of Things (IoT) \cite{Whitmore2015}, has witnessed exponential growth in recent years, evolving into an integral component of various application domains including healthcare, agriculture, manufacturing, transportation, and smart cities \cite{Sutikno2022}. The number of connected IoT devices is expected to reach 20 billion by 2030, a great escalation from the 8.4 billion recorded in 2017 \cite{Statista2019}. This proliferation shows not only the ubiquity of IoT but also the increasing complexity and diversity of its applications.

Central to the IoT ecosystem are various technologies and protocols that facilitate seamless communication between connected devices \cite{Lin2017}. Among these, the Constrained Application Protocol (CoAP) emerges as one of the main architectural components. As a lightweight and efficient web-like transfer protocol, CoAP is meticulously tailored for IoT devices constrained by limited resources \cite{Shelby2014}. Usually operated over UDP (User Datagram Protocol), CoAP ensures a streamlined and reliable data exchange, mitigating overhead and minimizing power consumption. These attributes make CoAP a good choice in applications demanding low latency, real-time interaction, and support for multicast communications, being pivotal in scenarios like sensor networks and real-time monitoring.

However, the use of the CoAP protocol in IoT applications poses a number of challenges, particularly in the case of cybersecurity. The intrinsic characteristics of CoAP, coupled with the constraints inherent to IoT devices and networks — such as limited computational power, memory, and energy resources — supposes a source of cybersecurity vulnerabilities \cite{Rizzardi2022}. Addressing these challenges demands the implementation of robust security mechanisms. While employing DTLS (Datagram Transport Layer Security) helps secure CoAP communications in those cases when it can be implemented \cite{9220033}, it is equally crucial to integrate monitoring and intrusion detection systems for timely detection and response to potential threats.

Many modern cybersecurity solutions for IoT infrastructures include the use of model-based Intrusion Detection Systems (IDS). These systems, leveraging artificial intelligence or statistical techniques, construct a baseline model of normal behavior in IoT environments, enabling them to identify deviations that could indicate attacks \cite{De2023}. This model-based approach is particularly advantageous in IoT settings, given the heterogeneity and complexity of IoT devices and their communication patterns. Traditional signature-based IDS approaches often fall short in such dynamic environments, as maintaining updated signatures for all possible attacks is a huge challenge \cite{Hassan2023}.

Focusing on the CoAP protocol, the application of model-based IDS is particularly promising. These systems can scrutinize communication patterns and detect anomalies indicative of a range of threats, such as Denial-of-Service (DoS) attacks, reconnaissance attacks, or unauthorized resource access. By analyzing metrics like message exchanges, response times, and payload sizes, model-based IDS can discern unusual behavior that may signal an attack, even in the absence of known attack signatures \cite{Granjal2018}.

Recent advancements in the field of machine learning offer new horizons for enhancing IoT security. Autoencoders, a deep learning technique, have been increasingly applied to build sophisticated anomaly detection systems for IoT \cite{ABUSITTA2023100656}. These autoencoders excel at acquiring a proficient representation of input data, a quality that proves to be very valuable in the creation of efficient network intrusion detection systems. \cite{Yao2023}. However, the existing research often overlooks or does not focus specifically on CoAP network communications or vulnerabilities, highlighting a critical gap this paper aims to address.

This paper presents a significant scientific contribution by advancing the application of autoencoders in the field of IoT cybersecurity. Specifically, it innovates by optimizing the use of the latent space within autoencoders to efficiently classify three different types of attacks and normal frames in an IoT environment dataset focused on the Constrained Application Protocol (CoAP). The exploration of the latent space not only aids in reducing the dimensionality of data, thereby enhancing processing efficiency, but also improves the accuracy and speed of anomaly detection in Intrusion Detection Systems (IDS). This dual enhancement—efficiency in data processing and effectiveness in anomaly detection—represents a substantial advancement in the capabilities of IDS within IoT networks, addressing critical security challenges in real-time environments. By refining the model to effectively utilize latent space, this work contributes to the development of more intelligent, scalable, and robust cybersecurity solutions for IoT systems.

The rest of the paper is organized as follows: Section 2 introduces a new dataset developed to target specific vulnerabilities of CoAP, leveraging a simulated IoT environment. The section describes the operational details of the CoAP protocol and the methodology of conducting controlled attacks to generate relevant data for anomaly detection. In section 3, the use of autoencoders for feature reduction in deep learning is explored, emphasizing the efficiency of autoencoders in compressing and reconstructing input data, with a specific focus on the case of CoAP vulnerabilities. Section 4 discusses the use of decision trees in machine learning, emphasizing their structured approach and interpretability for handling non-linear feature relationships and comparing this approach to ensemble methods like Random Forest and XGBoost, in the context of intrusion detection scenarios. The experimental setup, including data transformation and methodology, is described in section 5. Key processes such as feature selection, normalization, autoencoder configuration, and classification methods are covered. Next, the experimental results are presented, focusing on usual metrics across different classifiers and autoencoder configurations, revealing the superior performance of ensemble methods over decision trees, especially in scenarios with limited features, and discusses the implications of these findings for CoAP attack detection. Finally, section 6 gives some conclusions and highlight some possible future research directions.

\section{Methodology}

\subsection{Dataset}

Due to the significance of datasets for enhancing the detection capabilities of Intrusion Detection Systems (IDS) in IoT environments, recent years have seen the development of datasets specifically for IoT settings \cite{Dutta2022}. Among these, the most recent is the CoAP-IoT \cite{Vigoya2023} dataset, which employs a simulated environment with message modification attacks to generate anomalies. Additionally, there is the CoAP-DoS \cite{Mathews2022} dataset, focused on denial-of-service attacks. To improve upon existing datasets, a new one has been developed that targets the specific vulnerabilities of the CoAP protocol as defined in its RFC7252 \cite{Shelby2014}.


CoAP, an acronym for "Constrained Application Protocol," operates at the application layer of the OSI model and is designed to emulate the functionality of the HTTP protocol in devices with energy and capacity constraints, such as IoT devices. Its main advantage over HTTP is its lightweight design, CoAP is based on the UDP (User Datagram Protocol) for data transmission.

CoAP follows the client/server model, and adopts the RESTful API architecture for communication. Resources are identified by unique URLs and are managed in  similar to HTTP  with the GET, POST, PUT, and DELETE methods. It extends functionality with the “Observe” method \cite{Correia2016} which enhances client-server interaction by allowing devices to subscribe to resources and receive automatic updates.

An environment for the CoAP protocol has been created using a Node.js server with the "node-coap" library, which receives and stores temperature and humidity data through POST requests. The server also provides real-time updates to subscribing clients through the “observe” function, essential for continuous monitoring of environmental conditions. A DHT11 sensor connected to a NodeMCU board, programmed with "ESP-CoAP" \cite{ESP-COAP2021} is used to transmit environmental data. A JavaScript client on a Raspberry Pi 3 visualizes this data, using the "Observe" function for constant updates. In addition, Copper4Cr \cite{copper2022}  clients are used for efficient interaction with the CoAP server, including sending requests and receiving data. This set of tools allows simulating a realistic IoT environment and performing controlled attacks to identify vulnerabilities in the CoAP protocol, generating useful data for anomaly detection.



To assess the security of the CoAP protocol, a series of controlled attacks will be executed within the specially designed environment. The primary objective of these attacks is to compile a comprehensive dataset that will be instrumental in identifying and analyzing anomalies within the protocol. To capture this data effectively, a router equipped with OpenWRT is employed to monitor and record all the traffic traversing the local area network (LAN). This method guarantees the collection of an authentic dataset, which closely mirrors the scenarios that might occur in an IoT setting operating on the CoAP protocol when it is subjected to cyber threats. This dataset will serve as a critical resource for understanding the behavior of the CoAP protocol under duress and for enhancing the detection capabilities of Intrusion Detection Systems (IDS) in IoT environments.




  \textbf{Denial of service attack}: CoAP servers are designed to respond to requests with packets that can be significantly larger than the incoming requests. This is due to CoAP's block-wise transfer capability, which allows the transmission of data in adjustable block sizes. In some cases, especially during an attack, these blocks can be configured to be exceedingly small, leading to a potential vulnerability where CoAP clients can be overwhelmed by Denial of Service (DoS) attacks\cite{Thomas2017,Mathews2022}.

To exploit this vulnerability, an attacker can launch an amplification attack by spoofing the IP address of a legitimate CoAP client. This is done by manipulating the request packet to appear as if it originates from the victim's IP address, prompting the server to send the response to the victim instead of the attacker. The attacker, posing as a client using the Copper4Cr tool, tricks the server into responding to their requests, which are actually directed at the legitimate client.

\textbf{The Man-in-The-Middle (MitM)} attack exploits the CoAP protocol by intercepting the communication between a client and a server, a technique known as "sniffing"\cite{Fereidouni2023}. The attacker, positioned between the client and server, uses Kali Linux and the Ettercap  tool \cite{Arreaga2023} similar to the approach described in the MQTT section "4.2.2.3.2 Man-in-the-middle Attack." By employing Ettercap filters, which compile source files into binaries that Ettercap can execute, the attacker can manipulate the CoAP communication. They modify an existing filter for TCP and HTTP requests to target UDP requests of the CoAP protocol. Consequently, any POST or GET request made by the client to the "/temp" route is rerouted to the "/mitm" path on the attacker's machine. This causes the server to attempt to respond to a non-existent route, leaving the original client request unanswered, while the client remains unaware of the interference, believing it has communicated correctly.

\textbf{The cross-protocol attack} as detailed in RFC7252 under section 11.5 "Cross-Protocol Attacks," involves a method where the attacker sends DNS packets to the client while masquerading as the server's IP address. The client, expecting CoAP protocol communication, mistakenly interprets these DNS packets as legitimate CoAP messages due to their transmission over the same UDP protocol. This type of attack exploits the structural similarities between CoAP and other UDP-based protocols like DNS, making CoAP susceptible to such deceptive tactics.

The danger of this attack lies in its ability to bypass firewall settings that are specifically configured for CoAP traffic, thereby enabling the attacker to perform denial-of-service attacks or inject malicious code into various IoT system clients. To execute this attack, the attacker employs a node.js script equipped with the "native-dns-packet" library. This library provides the functionality to meticulously edit the parameters of a DNS packet. The attacker adjusts the bit length of each field to correspond with the content and port of a CoAP message. This manipulation is so precise that it alters the data transmitted from the temperature sensor using the "observe" function, effectively changing the reported temperature readings.
 

Attacks within the network are collected by capturing all traffic, resulting in three distinct PCAP files for each type of attack. Given the complexity of these PCAP files, which contain an extensive array of fields not pertinent to the detection of attacks on the CoAP protocol, a detailed dissection is performed. This process involves isolating the frames of all generated traffic and extracting the common fields across these frames. Essential information such as system timestamps, relative capture times, and all CoAP protocol-related fields are preserved. Each frame is then labeled according to its timestamp at the moment of capture, designating whether it is part of an attack scenario
 
In conclusion, the dataset encompasses a trio of CSV files, each featuring a uniform number of fields (totaling 68). These fields have been meticulously curated utilizing the Wireshark Display Filter Reference [1]. This selection includes fields present at the frame level, universally applicable across all frames, alongside every field pertaining to the Restricted Application Protocol (CoAP). Additionally, a bespoke field labeled 'type' is incorporated to classify whether the frames are subject to an attack. Each file within the dataset is dedicated to documenting a specific category of attack.

    \begin{itemize}
    \item CoAP\_Cross\_protocol.CSV has 62,943 rows, one for each frame, with 60453 frames of normal traffic and 2490 frames of traffic under attack.
    \item CoAP\_MitM.CSV consisting of 24684 rows, one for each frame, with 21222 frames of normal traffic and 3462 frames of traffic under attack.
    \item CoAP\_DoS.CSV has 30319 rows, one for each frame, with 21269 frames of normal traffic and 9050 frames of traffic under attack
    \end{itemize}


\subsection{Feature reduction: Autoencoder}

Common class of deep learning techniques called autoencoders is made to learn a compact deep representation of data \cite{Pumsirirat2018}. They feature a symmetrical architecture, with the same number of neurons in the input and output layers. Compared to the input and output layers, there are less neurons in the middle hidden layer, sometimes referred to as the latent space.


While the decoder component of the autoencoder stretches from the latent space to the output layer, the encoder section goes from the input layer to the latent space. Reconstructing the input data is an autoencoder's main goal. The autoencoder efficiently compresses the input data down to the number of neurons existing in the latent space, provided that the data can be reliably recovered following the encoding and decoding procedures.


Understanding the behavior of CoAP vulnerabilities in this study will be greatly aided by the number of neurons in the latent space.

\subsection{Classification}\label{classification}
\subsubsection{Decision trees}

In machine learning research, decision trees \cite{kingsford2008decision} have been extensively acknowledged and employed. These trees exemplify workflows or procedures via tree configurations, presenting a methodical and lucid strategy for addressing challenges. Within the tree, each internal node signifies an examination of a particular attribute, with branches denoting test outcomes, and the terminal leaves designating class annotations. As a result, the trajectory from the root to a leaf delineates a classification directive, enabling unambiguous interpretation and determinations.
\\

Authors inclination towards decision trees as the research model is underpinned by multiple considerations. Primarily, the pronounced interpretability of decision trees is pivotal in sectors where model lucidity and intelligibility are paramount. The accessible nature of decision trees allows professionals and interested parties to elucidate and authenticate the decision-making paradigm. Such transparency amplifies confidence in the model's prognostications and aids in discerning pivotal attributes or decision trajectories.
\\

Additionally, decision trees exhibit proficiency in delineating non-linear associations and interplays between features. Their tiered architecture facilitates discernment of multifaceted decision boundaries, proving advantageous when data distribution manifests sophisticated configurations. The facility of decision trees to engage with diverse data categories, spanning both categorical and numerical features, bolsters their broad-spectrum applicability across varied fields.

\subsubsection{Ensemble methods}
On the other hand, ensemble methods have been also selected to evaluate their performance with respect decision trees.
Ensemble methods have gained prominence in recent years due to their remarkable ability to enhance predictive performance and model robustness \cite{TR2023100247} \cite{ELMEZUGHI2023e19685}. Unlike standalone decision trees, which are prone to overfitting and can be highly sensitive to the training data, ensemble methods combine multiple models to produce a more accurate and stable prediction. By aggregating the outputs of individual decision trees or other base models, ensemble methods can effectively mitigate the shortcomings of individual models. They harness the wisdom of the crowd, leveraging diverse perspectives and learning from different aspects of the data to make more reliable predictions. In essence, ensemble methods provide a valuable tool for improving model generalization, reducing variance, and achieving superior performance in various machine learning tasks.
\\
In this case, Extreme Gradient Boosting (XGBoost) and Random Forest have been used.
\\
\textbf{XGBoost} is a cutting-edge machine learning algorithm that has emerged as a formidable tool for predictive modeling and classification tasks \cite{ChenG16}. It stands out for its exceptional predictive accuracy and speed, making it a preferred choice in both research and industry applications. XGBoost employs a gradient boosting framework, which sequentially builds an ensemble of decision trees, each one correcting the errors of its predecessor. What sets XGBoost apart is its optimization techniques, such as regularization, parallel processing, and tree pruning, which enhance its efficiency and make it highly resistant to overfitting.
\\
\textbf{Random Forest (RF)} is a widely adopted ensemble learning technique that has demonstrated its effectiveness in various machine learning applications \cite{Breiman2001}. This method operates by constructing multiple decision trees during the training phase and combines their predictions through a process known as bagging (Bootstrap Aggregating). By introducing randomness in the tree construction process, Random Forest reduces the risk of overfitting and enhances the model's generalization capabilities. Additionally, it provides a valuable feature selection mechanism and an estimate of feature importance, making it a versatile tool for both classification and regression tasks. Its robustness, ease of use, and ability to handle high-dimensional data have made Random Forest a popular choice in the machine learning community \cite{SUN2024121549} \cite{JOSSO2023105671} \cite{JIANG2023145}.
\\
Random Forest and XGBoost are both ensemble learning techniques based on trees, but they diverge in several crucial aspects. Random Forest constructs an ensemble of decision trees independently, relying on bootstrap sampling and feature randomness for diversity. In contrast, XGBoost employs a gradient boosting framework, sequentially building decision trees to correct the errors of previous ones. XGBoost's focus on optimization and gradient descent techniques often leads to superior predictive performance, especially when finely tuned, making it a preferred choice for high-accuracy tasks. Additionally, XGBoost offers advanced solutions for handling imbalanced data and typically exhibits faster training times, while Random Forest is appreciated for its simplicity, robustness, and ease of use.


\section{Experiments}

This section offers the experimental arrangement, accompanied by an examination of the conclusive outcomes. Furthermore, it includes a succinct analysis of the ramifications associated with these results.

\subsection{Experimental setup}


The initial phase executed involves the transformation of the MAC address. This process begins with the alteration of the MAC address by eliminating the colons, followed by its conversion into numerical values. 
Following these set of categorical fields:

\begin{itemize}
    \item coap.opt.ctype
    \item coap.opt.desc
    \item coap.opt.name 
    \item coap.opt.uri\_path
    \item coap.payloa\_ddesc
    \item coap.token 
\end{itemize}
have been transformed into numerical values utilizing the OneHotEncoder function from the scikit-learn library. Moreover, a comprehensive analysis of 84 features was conducted to identify and eliminate those lacking substantial representational diversity. This includes features with constant or null values. Ultimately, a refined set of 52 features was selected for use. This set comprises features derived both from the original columns and from the columns generated through the OneHotEncoding of the six categorical columns.

After these transformations, it is imperative to normalize the data to facilitate effective training via machine learning algorithms. In this specific scenario, textit{MinMax} normalization was employed. This technique constrains the value of each characteristic to fall within a range from 0 to 1. Following the normalization process, an exploratory study using an auto-encoder was undertaken to perform feature reduction.


To achieve precise classification of cyber attacks, the researchers conducted an analysis focusing on the importance of the latent space within the Autoencoder (AE). The AE was configured to use 'relu' activation functions for all layers, with the exception of the latent space where a linear activation function was implemented. The training of the AE extended over 50 epochs, employing a batch size of 50. Mean Square Error (MSE) was utilized as the loss function, and 'adam' was selected as the optimizer for this process.


The autoencoder designed for this study comprises seven layers, each with a configuration of 48-35-28-N-28-35-48 neurons, where N varies within the range of [1,27]. Post the training phase of the network, the dataset underwent an encoding process to diminish its feature count from 48 to \textit{n}, with \textit{n} spanning from 1 to 27. This dimensionally reduced dataset was subsequently utilized in the classification stage of the study.

As mentioned in Section \ref{classification}, the classification methods chosen were decision trees and two ensemble techniques: Random Forest and XGBoost. For decision trees, a hyperparameter search was conducted using \textit{GridSearchCV}. This search involved cross-validation, with the maximum number of leaves and the criteria (Gini or entropy) varied during the search. Similarly, for Random Forest, we conducted a hyperparameter search with the number of estimators and the criteria. For XGBoost, a hyperparameter search was performed over the number of estimators and the maximum depth of the trees


To address the issue of class imbalance, a class weighting strategy has been implemented. 
In this proposed method, the \textit{class-weight} function is utilized to automatically allocate appropriate weights to each class based on its frequency within the dataset.

This enables the model to equally learn from both minority and majority classes, thereby effectively mitigating the inherent bias often observed in imbalanced datasets. By seamlessly integrating the class weighting technique into the classification pipeline, and capitalizing on the extensive functionalities provided by the scikit-learn library, the overall performance of the model is substantially enhanced. As a result, the model becomes more adept at making informed decisions for both underrepresented and overrepresented classes.

\begin{figure}[!h]
    \centering
    \includegraphics[width=13cm]{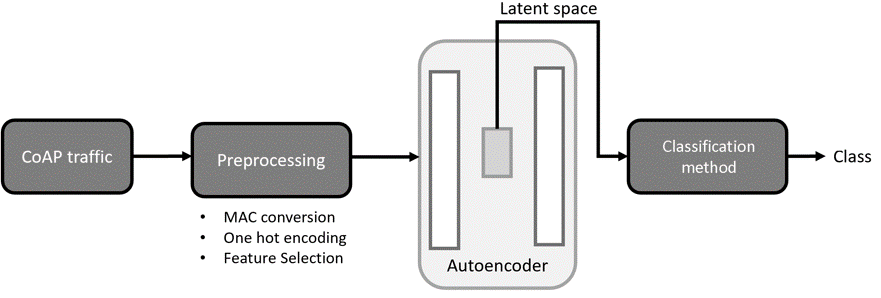}
    \caption{Schema with the complete flow of the experiments}
    \label{fig.schema}
\end{figure}

The complete experiment flow can be shown better in the figure
\ref{fig.schema}

Different experiments have been carried out with Decision Trees and the two ensemble methods evaluated XGBoost and Random Forest using the latent space with different feature sizes.


The metrics of Precision, Recall, and F-Score are computed individually for each class has been implemented for all categories. A weighted average is also defined by factoring in the proportion of data belonging to each respective class.

\subsection{Results and discussion}

Precision, Recall, and F-Score was systematically undertaken for each classification category. This evaluation encompassed every designated attack class, additionally incorporating a weighted average calculation that took into account the proportional representation of each class within the dataset. To facilitate a comprehensive analysis, the mean results corresponding to varying sizes of autoencoders have been meticulously tabulated and can be referenced in Table \ref{results}. This table provides an insightful comparative perspective on the performance metrics across different autoencoder configurations.

The results highlight a consistent pattern of improvement in precision, recall, and F1-score metrics when ensemble techniques such as Random Forest (RF) and XGBoost (XGB) are employed compared to using standalone Decision Trees (DT). Random Forest (RF) emerges as a standout performer, especially in scenarios with a limited number of features. Even with a sparse feature set, RF achieves precision, recall, and F1-score values that are notably higher than those of standalone Decision Trees. This emphasises the advantage of the ensemble approach, which combines multiple decision trees to produce a more robust and accurate predictive model. Random Forest algorithm needs only two features in the latent space to achieve a 0.99 in all the metrics while decision trees needs the double and XGBoost three features.

The results clearly illustrate that ensemble techniques, represented here by RF and XGB, have a distinct advantage over standalone Decision Trees. They not only offer improved performance but also showcase the capacity to make better use of available features. This underscores the importance of leveraging ensemble methods when working with decision trees, particularly when confronted with datasets characterized by limited feature dimensions. 
\begin{table}[]

\centering
\caption{Precision, recall and f-score for all the different latent space sizes using decision trees (DT) and the ensemble methods Random Forest (RF) and XGBoost (XGB). Highlighted the results higher than 0.99.}
\label{results}
\begin{tabular}{l|lll|lll|lll|}
\cline{2-10}
\textbf{}                               & \multicolumn{3}{c|}{\textbf{PRECISION}}                                                                                                    & \multicolumn{3}{c|}{\textbf{RECALL}}                                                                                                       & \multicolumn{3}{c|}{\textbf{F-SCORE}}                                                                                                      \\ \hline
\multicolumn{1}{|l|}{\textbf{Features}} & \multicolumn{1}{l|}{\textbf{DT}}                    & \multicolumn{1}{l|}{\textbf{RF}}                    & \textbf{XGB}                   & \multicolumn{1}{l|}{\textbf{DT}}                    & \multicolumn{1}{l|}{\textbf{RF}}                    & \textbf{XGB}                   & \multicolumn{1}{l|}{\textbf{DT}}                    & \multicolumn{1}{l|}{\textbf{RF}}                    & \textbf{XGB}                   \\ \hline
\multicolumn{1}{|l|}{\textbf{1}}        & \multicolumn{1}{l|}{0.9415}                         & \multicolumn{1}{l|}{0.9553}                         & 0.9365                         & \multicolumn{1}{l|}{0.8023}                         & \multicolumn{1}{l|}{0.9554}                         & 0.9438                         & \multicolumn{1}{l|}{0.8535}                         & \multicolumn{1}{l|}{0.9552}                         & 0.9340                         \\ \hline
\multicolumn{1}{|l|}{\textbf{2}}        & \multicolumn{1}{l|}{0.9654}                         & \multicolumn{1}{l|}{\cellcolor[HTML]{FFCCC9}0.9919} & 0.9878                         & \multicolumn{1}{l|}{0.9443}                         & \multicolumn{1}{l|}{\cellcolor[HTML]{FFCCC9}0.9920} & 0.9880                         & \multicolumn{1}{l|}{0.9511}                         & \multicolumn{1}{l|}{\cellcolor[HTML]{FFCCC9}0.9919} & 0.9878                         \\ \hline
\multicolumn{1}{|l|}{\textbf{3}}        & \multicolumn{1}{l|}{0.9552}                         & \multicolumn{1}{l|}{\cellcolor[HTML]{FFCCC9}0.9934} & \cellcolor[HTML]{FFCCC9}0.9938 & \multicolumn{1}{l|}{0.9142}                         & \multicolumn{1}{l|}{\cellcolor[HTML]{FFCCC9}0.9936} & \cellcolor[HTML]{FFCCC9}0.9938 & \multicolumn{1}{l|}{0.9278}                         & \multicolumn{1}{l|}{\cellcolor[HTML]{FFCCC9}0.9936} & \cellcolor[HTML]{FFCCC9}0.9937 \\ \hline
\multicolumn{1}{|l|}{\textbf{4}}        & \multicolumn{1}{l|}{\cellcolor[HTML]{FFCCC9}0.9997} & \multicolumn{1}{l|}{\cellcolor[HTML]{FFCCC9}0.9996} & \cellcolor[HTML]{FFCCC9}0.9997 & \multicolumn{1}{l|}{\cellcolor[HTML]{FFCCC9}0.9997} & \multicolumn{1}{l|}{\cellcolor[HTML]{FFCCC9}0.9996} & \cellcolor[HTML]{FFCCC9}0.9997 & \multicolumn{1}{l|}{\cellcolor[HTML]{FFCCC9}0.9997} & \multicolumn{1}{l|}{\cellcolor[HTML]{FFCCC9}0.9996} & \cellcolor[HTML]{FFCCC9}0.9996 \\ \hline
\multicolumn{1}{|l|}{\textbf{5}}        & \multicolumn{1}{l|}{\cellcolor[HTML]{FFCCC9}0.9997} & \multicolumn{1}{l|}{\cellcolor[HTML]{FFCCC9}0.9997} & \cellcolor[HTML]{FFCCC9}0.9997 & \multicolumn{1}{l|}{\cellcolor[HTML]{FFCCC9}0.9997} & \multicolumn{1}{l|}{\cellcolor[HTML]{FFCCC9}0.9997} & \cellcolor[HTML]{FFCCC9}0.9997 & \multicolumn{1}{l|}{\cellcolor[HTML]{FFCCC9}0.9997} & \multicolumn{1}{l|}{\cellcolor[HTML]{FFCCC9}0.9997} & \cellcolor[HTML]{FFCCC9}0.9997 \\ \hline
\multicolumn{1}{|l|}{\textbf{6}}        & \multicolumn{1}{l|}{\cellcolor[HTML]{FFCCC9}0.9997} & \multicolumn{1}{l|}{\cellcolor[HTML]{FFCCC9}0.9998} & \cellcolor[HTML]{FFCCC9}0.9999 & \multicolumn{1}{l|}{\cellcolor[HTML]{FFCCC9}0.9997} & \multicolumn{1}{l|}{\cellcolor[HTML]{FFCCC9}0.9998} & \cellcolor[HTML]{FFCCC9}0.9999 & \multicolumn{1}{l|}{\cellcolor[HTML]{FFCCC9}0.9997} & \multicolumn{1}{l|}{\cellcolor[HTML]{FFCCC9}0.9998} & \cellcolor[HTML]{FFCCC9}0.9999 \\ \hline
\multicolumn{1}{|l|}{\textbf{7}}        & \multicolumn{1}{l|}{\cellcolor[HTML]{FFCCC9}0.9995} & \multicolumn{1}{l|}{\cellcolor[HTML]{FFCCC9}0.9998} & \cellcolor[HTML]{FFCCC9}0.9997 & \multicolumn{1}{l|}{\cellcolor[HTML]{FFCCC9}0.9995} & \multicolumn{1}{l|}{\cellcolor[HTML]{FFCCC9}0.9998} & \cellcolor[HTML]{FFCCC9}0.9997 & \multicolumn{1}{l|}{\cellcolor[HTML]{FFCCC9}0.9995} & \multicolumn{1}{l|}{\cellcolor[HTML]{FFCCC9}0.9998} & \cellcolor[HTML]{FFCCC9}0.9997 \\ \hline
\multicolumn{1}{|l|}{\textbf{8}}        & \multicolumn{1}{l|}{\cellcolor[HTML]{FFCCC9}0.9997} & \multicolumn{1}{l|}{\cellcolor[HTML]{FFCCC9}0.9997} & \cellcolor[HTML]{FFCCC9}0.9998 & \multicolumn{1}{l|}{\cellcolor[HTML]{FFCCC9}0.9997} & \multicolumn{1}{l|}{\cellcolor[HTML]{FFCCC9}0.9997} & \cellcolor[HTML]{FFCCC9}0.9998 & \multicolumn{1}{l|}{\cellcolor[HTML]{FFCCC9}0.9997} & \multicolumn{1}{l|}{\cellcolor[HTML]{FFCCC9}0.9997} & \cellcolor[HTML]{FFCCC9}0.9998 \\ \hline
\multicolumn{1}{|l|}{\textbf{9}}        & \multicolumn{1}{l|}{\cellcolor[HTML]{FFCCC9}0.9985} & \multicolumn{1}{l|}{\cellcolor[HTML]{FFCCC9}0.9998} & \cellcolor[HTML]{FFCCC9}0.9998 & \multicolumn{1}{l|}{\cellcolor[HTML]{FFCCC9}0.9981} & \multicolumn{1}{l|}{\cellcolor[HTML]{FFCCC9}0.9998} & \cellcolor[HTML]{FFCCC9}0.9998 & \multicolumn{1}{l|}{\cellcolor[HTML]{FFCCC9}0.9982} & \multicolumn{1}{l|}{\cellcolor[HTML]{FFCCC9}0.9998} & \cellcolor[HTML]{FFCCC9}0.9998 \\ \hline
\multicolumn{1}{|l|}{\textbf{10}}       & \multicolumn{1}{l|}{\cellcolor[HTML]{FFCCC9}0.9997} & \multicolumn{1}{l|}{\cellcolor[HTML]{FFCCC9}0.9998} & \cellcolor[HTML]{FFCCC9}0.9998 & \multicolumn{1}{l|}{\cellcolor[HTML]{FFCCC9}0.9997} & \multicolumn{1}{l|}{\cellcolor[HTML]{FFCCC9}0.9998} & \cellcolor[HTML]{FFCCC9}0.9998 & \multicolumn{1}{l|}{\cellcolor[HTML]{FFCCC9}0.9997} & \multicolumn{1}{l|}{\cellcolor[HTML]{FFCCC9}0.9998} & \cellcolor[HTML]{FFCCC9}0.9998 \\ \hline
\multicolumn{1}{|l|}{\textbf{11}}       & \multicolumn{1}{l|}{\cellcolor[HTML]{FFCCC9}0.9999} & \multicolumn{1}{l|}{\cellcolor[HTML]{FFCCC9}0.9999} & \cellcolor[HTML]{FFCCC9}1.0000 & \multicolumn{1}{l|}{\cellcolor[HTML]{FFCCC9}0.9999} & \multicolumn{1}{l|}{\cellcolor[HTML]{FFCCC9}0.9999} & \cellcolor[HTML]{FFCCC9}1.0000 & \multicolumn{1}{l|}{\cellcolor[HTML]{FFCCC9}0.9999} & \multicolumn{1}{l|}{\cellcolor[HTML]{FFCCC9}0.9999} & \cellcolor[HTML]{FFCCC9}1.0000 \\ \hline
\multicolumn{1}{|l|}{\textbf{12}}       & \multicolumn{1}{l|}{\cellcolor[HTML]{FFCCC9}0.9995} & \multicolumn{1}{l|}{\cellcolor[HTML]{FFCCC9}0.9998} & \cellcolor[HTML]{FFCCC9}0.9998 & \multicolumn{1}{l|}{\cellcolor[HTML]{FFCCC9}0.9995} & \multicolumn{1}{l|}{\cellcolor[HTML]{FFCCC9}0.9998} & \cellcolor[HTML]{FFCCC9}0.9998 & \multicolumn{1}{l|}{\cellcolor[HTML]{FFCCC9}0.9995} & \multicolumn{1}{l|}{\cellcolor[HTML]{FFCCC9}0.9998} & \cellcolor[HTML]{FFCCC9}0.9998 \\ \hline
\multicolumn{1}{|l|}{\textbf{13}}       & \multicolumn{1}{l|}{\cellcolor[HTML]{FFCCC9}0.9998} & \multicolumn{1}{l|}{\cellcolor[HTML]{FFCCC9}0.9999} & \cellcolor[HTML]{FFCCC9}1.0000 & \multicolumn{1}{l|}{\cellcolor[HTML]{FFCCC9}0.9998} & \multicolumn{1}{l|}{\cellcolor[HTML]{FFCCC9}0.9999} & \cellcolor[HTML]{FFCCC9}1.0000 & \multicolumn{1}{l|}{\cellcolor[HTML]{FFCCC9}0.9998} & \multicolumn{1}{l|}{\cellcolor[HTML]{FFCCC9}0.9999} & \cellcolor[HTML]{FFCCC9}1.0000 \\ \hline
\multicolumn{1}{|l|}{\textbf{14}}       & \multicolumn{1}{l|}{\cellcolor[HTML]{FFCCC9}0.9997} & \multicolumn{1}{l|}{\cellcolor[HTML]{FFCCC9}0.9997} & \cellcolor[HTML]{FFCCC9}0.9997 & \multicolumn{1}{l|}{\cellcolor[HTML]{FFCCC9}0.9997} & \multicolumn{1}{l|}{\cellcolor[HTML]{FFCCC9}0.9997} & \cellcolor[HTML]{FFCCC9}0.9997 & \multicolumn{1}{l|}{\cellcolor[HTML]{FFCCC9}0.9997} & \multicolumn{1}{l|}{\cellcolor[HTML]{FFCCC9}0.9997} & \cellcolor[HTML]{FFCCC9}0.9997 \\ \hline
\multicolumn{1}{|l|}{\textbf{15}}       & \multicolumn{1}{l|}{\cellcolor[HTML]{FFCCC9}0.9998} & \multicolumn{1}{l|}{\cellcolor[HTML]{FFCCC9}0.9997} & \cellcolor[HTML]{FFCCC9}0.9998 & \multicolumn{1}{l|}{\cellcolor[HTML]{FFCCC9}0.9998} & \multicolumn{1}{l|}{\cellcolor[HTML]{FFCCC9}0.9997} & \cellcolor[HTML]{FFCCC9}0.9998 & \multicolumn{1}{l|}{\cellcolor[HTML]{FFCCC9}0.9998} & \multicolumn{1}{l|}{\cellcolor[HTML]{FFCCC9}0.9997} & \cellcolor[HTML]{FFCCC9}0.9998 \\ \hline
\multicolumn{1}{|l|}{\textbf{16}}       & \multicolumn{1}{l|}{\cellcolor[HTML]{FFCCC9}0.9997} & \multicolumn{1}{l|}{\cellcolor[HTML]{FFCCC9}0.9999} & \cellcolor[HTML]{FFCCC9}0.9999 & \multicolumn{1}{l|}{\cellcolor[HTML]{FFCCC9}0.9997} & \multicolumn{1}{l|}{\cellcolor[HTML]{FFCCC9}0.9999} & \cellcolor[HTML]{FFCCC9}0.9999 & \multicolumn{1}{l|}{\cellcolor[HTML]{FFCCC9}0.9997} & \multicolumn{1}{l|}{\cellcolor[HTML]{FFCCC9}0.9999} & \cellcolor[HTML]{FFCCC9}0.9999 \\ \hline
\multicolumn{1}{|l|}{\textbf{17}}       & \multicolumn{1}{l|}{\cellcolor[HTML]{FFCCC9}0.9997} & \multicolumn{1}{l|}{\cellcolor[HTML]{FFCCC9}0.9999} & \cellcolor[HTML]{FFCCC9}0.9999 & \multicolumn{1}{l|}{\cellcolor[HTML]{FFCCC9}0.9997} & \multicolumn{1}{l|}{\cellcolor[HTML]{FFCCC9}0.9999} & \cellcolor[HTML]{FFCCC9}0.9999 & \multicolumn{1}{l|}{\cellcolor[HTML]{FFCCC9}0.9997} & \multicolumn{1}{l|}{\cellcolor[HTML]{FFCCC9}0.9999} & \cellcolor[HTML]{FFCCC9}0.9999 \\ \hline
\multicolumn{1}{|l|}{\textbf{18}}       & \multicolumn{1}{l|}{\cellcolor[HTML]{FFCCC9}0.9997} & \multicolumn{1}{l|}{\cellcolor[HTML]{FFCCC9}0.9997} & \cellcolor[HTML]{FFCCC9}0.9998 & \multicolumn{1}{l|}{\cellcolor[HTML]{FFCCC9}0.9997} & \multicolumn{1}{l|}{\cellcolor[HTML]{FFCCC9}0.9997} & \cellcolor[HTML]{FFCCC9}0.9998 & \multicolumn{1}{l|}{\cellcolor[HTML]{FFCCC9}0.9997} & \multicolumn{1}{l|}{\cellcolor[HTML]{FFCCC9}0.9997} & \cellcolor[HTML]{FFCCC9}0.9998 \\ \hline
\multicolumn{1}{|l|}{\textbf{19}}       & \multicolumn{1}{l|}{\cellcolor[HTML]{FFCCC9}0.9997} & \multicolumn{1}{l|}{\cellcolor[HTML]{FFCCC9}0.9999} & \cellcolor[HTML]{FFCCC9}0.9998 & \multicolumn{1}{l|}{\cellcolor[HTML]{FFCCC9}0.9997} & \multicolumn{1}{l|}{\cellcolor[HTML]{FFCCC9}0.9999} & \cellcolor[HTML]{FFCCC9}0.9998 & \multicolumn{1}{l|}{\cellcolor[HTML]{FFCCC9}0.9997} & \multicolumn{1}{l|}{\cellcolor[HTML]{FFCCC9}0.9999} & \cellcolor[HTML]{FFCCC9}0.9998 \\ \hline
\multicolumn{1}{|l|}{\textbf{20}}       & \multicolumn{1}{l|}{\cellcolor[HTML]{FFCCC9}0.9992} & \multicolumn{1}{l|}{\cellcolor[HTML]{FFCCC9}0.9998} & \cellcolor[HTML]{FFCCC9}0.9998 & \multicolumn{1}{l|}{\cellcolor[HTML]{FFCCC9}0.9992} & \multicolumn{1}{l|}{\cellcolor[HTML]{FFCCC9}0.9998} & \cellcolor[HTML]{FFCCC9}0.9998 & \multicolumn{1}{l|}{\cellcolor[HTML]{FFCCC9}0.9992} & \multicolumn{1}{l|}{\cellcolor[HTML]{FFCCC9}0.9998} & \cellcolor[HTML]{FFCCC9}0.9998 \\ \hline
\multicolumn{1}{|l|}{\textbf{21}}       & \multicolumn{1}{l|}{\cellcolor[HTML]{FFCCC9}0.9998} & \multicolumn{1}{l|}{\cellcolor[HTML]{FFCCC9}0.9999} & \cellcolor[HTML]{FFCCC9}1.0000 & \multicolumn{1}{l|}{\cellcolor[HTML]{FFCCC9}0.9998} & \multicolumn{1}{l|}{\cellcolor[HTML]{FFCCC9}0.9999} & \cellcolor[HTML]{FFCCC9}1.0000 & \multicolumn{1}{l|}{\cellcolor[HTML]{FFCCC9}0.9998} & \multicolumn{1}{l|}{\cellcolor[HTML]{FFCCC9}0.9999} & \cellcolor[HTML]{FFCCC9}1.0000 \\ \hline
\multicolumn{1}{|l|}{\textbf{22}}       & \multicolumn{1}{l|}{\cellcolor[HTML]{FFCCC9}0.9797} & \multicolumn{1}{l|}{\cellcolor[HTML]{FFCCC9}0.9981} & \cellcolor[HTML]{FFCCC9}0.9991 & \multicolumn{1}{l|}{\cellcolor[HTML]{FFCCC9}0.9746} & \multicolumn{1}{l|}{\cellcolor[HTML]{FFCCC9}0.9982} & \cellcolor[HTML]{FFCCC9}0.9991 & \multicolumn{1}{l|}{\cellcolor[HTML]{FFCCC9}0.9761} & \multicolumn{1}{l|}{\cellcolor[HTML]{FFCCC9}0.9982} & \cellcolor[HTML]{FFCCC9}0.9991 \\ \hline
\multicolumn{1}{|l|}{\textbf{23}}       & \multicolumn{1}{l|}{\cellcolor[HTML]{FFCCC9}0.9999} & \multicolumn{1}{l|}{\cellcolor[HTML]{FFCCC9}0.9999} & \cellcolor[HTML]{FFCCC9}0.9999 & \multicolumn{1}{l|}{\cellcolor[HTML]{FFCCC9}0.9999} & \multicolumn{1}{l|}{\cellcolor[HTML]{FFCCC9}0.9999} & \cellcolor[HTML]{FFCCC9}0.9999 & \multicolumn{1}{l|}{\cellcolor[HTML]{FFCCC9}0.9999} & \multicolumn{1}{l|}{\cellcolor[HTML]{FFCCC9}0.9999} & \cellcolor[HTML]{FFCCC9}0.9999 \\ \hline
\multicolumn{1}{|l|}{\textbf{24}}       & \multicolumn{1}{l|}{\cellcolor[HTML]{FFCCC9}0.9995} & \multicolumn{1}{l|}{\cellcolor[HTML]{FFCCC9}0.9997} & \cellcolor[HTML]{FFCCC9}0.9997 & \multicolumn{1}{l|}{\cellcolor[HTML]{FFCCC9}0.9995} & \multicolumn{1}{l|}{\cellcolor[HTML]{FFCCC9}0.9997} & \cellcolor[HTML]{FFCCC9}0.9997 & \multicolumn{1}{l|}{\cellcolor[HTML]{FFCCC9}0.9995} & \multicolumn{1}{l|}{\cellcolor[HTML]{FFCCC9}0.9997} & \cellcolor[HTML]{FFCCC9}0.9997 \\ \hline
\multicolumn{1}{|l|}{\textbf{25}}       & \multicolumn{1}{l|}{\cellcolor[HTML]{FFCCC9}0.9997} & \multicolumn{1}{l|}{\cellcolor[HTML]{FFCCC9}0.9999} & \cellcolor[HTML]{FFCCC9}0.9998 & \multicolumn{1}{l|}{\cellcolor[HTML]{FFCCC9}0.9997} & \multicolumn{1}{l|}{\cellcolor[HTML]{FFCCC9}0.9999} & \cellcolor[HTML]{FFCCC9}0.9998 & \multicolumn{1}{l|}{\cellcolor[HTML]{FFCCC9}0.9997} & \multicolumn{1}{l|}{\cellcolor[HTML]{FFCCC9}0.9999} & \cellcolor[HTML]{FFCCC9}0.9998 \\ \hline
\multicolumn{1}{|l|}{\textbf{26}}       & \multicolumn{1}{l|}{\cellcolor[HTML]{FFCCC9}0.9998} & \multicolumn{1}{l|}{\cellcolor[HTML]{FFCCC9}0.9999} & \cellcolor[HTML]{FFCCC9}0.9999 & \multicolumn{1}{l|}{\cellcolor[HTML]{FFCCC9}0.9998} & \multicolumn{1}{l|}{\cellcolor[HTML]{FFCCC9}0.9999} & \cellcolor[HTML]{FFCCC9}0.9999 & \multicolumn{1}{l|}{\cellcolor[HTML]{FFCCC9}0.9998} & \multicolumn{1}{l|}{\cellcolor[HTML]{FFCCC9}0.9999} & \cellcolor[HTML]{FFCCC9}0.9999 \\ \hline
\multicolumn{1}{|l|}{\textbf{27}}       & \multicolumn{1}{l|}{\cellcolor[HTML]{FFCCC9}0.9996} & \multicolumn{1}{l|}{\cellcolor[HTML]{FFCCC9}0.9999} & \cellcolor[HTML]{FFCCC9}0.9999 & \multicolumn{1}{l|}{\cellcolor[HTML]{FFCCC9}0.9996} & \multicolumn{1}{l|}{\cellcolor[HTML]{FFCCC9}0.9999} & \cellcolor[HTML]{FFCCC9}0.9999 & \multicolumn{1}{l|}{\cellcolor[HTML]{FFCCC9}0.9996} & \multicolumn{1}{l|}{\cellcolor[HTML]{FFCCC9}0.9999} & \cellcolor[HTML]{FFCCC9}0.9999 \\ \hline
\end{tabular}
\end{table}

Our analysis reveals that XGBoost  obtains inferior results compared to Random Forest in almost all scenarios. XGBoost's boosting process, while capable of reducing bias in complex datasets, may also render it susceptible to overfitting, especially in cases with limited training data or inadequate regularization. Furthermore, The inherent robustness of Random Forest can lead to more stable performance when dealing with small or noisy data sets.

\begin{figure}[!h]
    \centering
    \includegraphics[width=8.5cm]{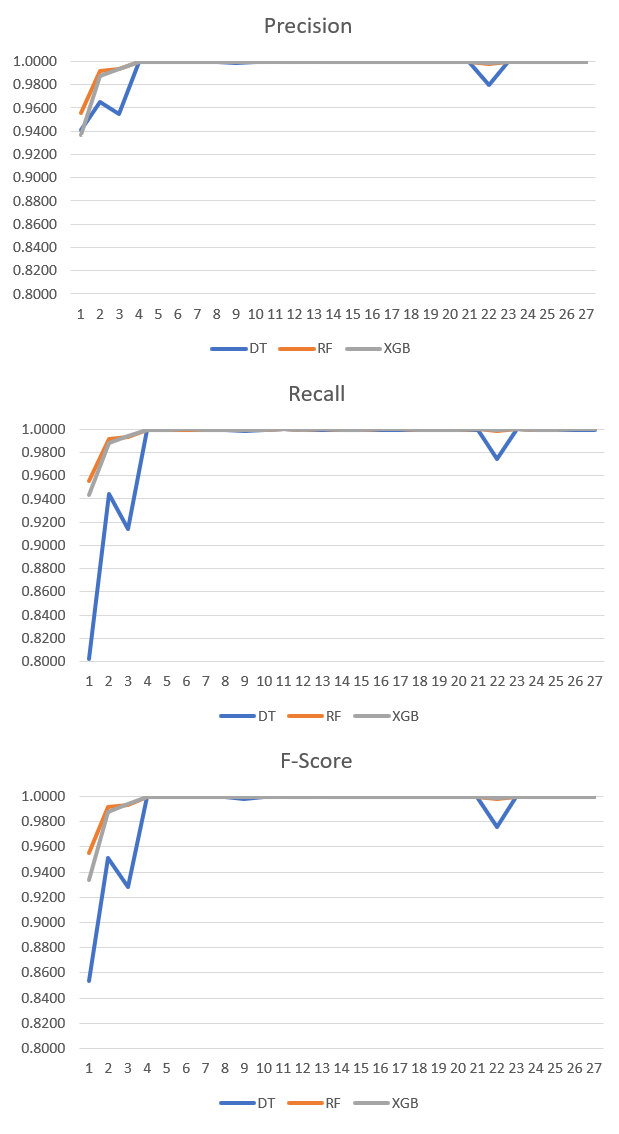}
    \caption{Precision, recall and F-Score class attack}
    \label{fig.resultadosClases}
\end{figure}

 The mean values for each of the latent spaces can be seen in figure \ref{fig.resultadosClases}, therefore chart displays the weighted average results of Precision, Recall, and F-Score for different numbers of neurons in the latent space for all the classifiers evaluated. The X-axis shows the number of neurons, while the Y-axis shows the corresponding value of the metric. 

All three methods exhibit similar behavior when four or more dimensions of the latent space are used, where precision, recall, and F-score approach near-perfect values. However, a noteworthy observation is how the ensemble methods consistently maintain these high-performance results for 22 dimensions of latent space, while the decision tree experiences a significant drop in performance. This decrease can likely be attributed to the autoencoder's coding process, which ensemble techniques effectively mitigate. However, the most interesting aspect arises when considering the smaller number of latent spatial dimensions. At this time, ensemble methods demonstrate their best performance by quickly achieving optimal classification performance with only two dimensions. This ability to excel with minimal latent space dimensions is of utmost importance for improving overall model efficiency and resource utilization, marking a significant advantage of ensemble techniques in practical applications.

\section{Conclusions and future works}

This article offers a comprehensive analysis of the impact of feature reduction using autoencoders on IoT CoAP attacks, and the influence of classical and ensemble classifiers on predictive outcomes. The primary aim is to determine the minimum number of features required for characterizing CoAP packets, facilitating the deployment of a real-time IoT attack detection system.

Through varying the dimension of the latent space in an autoencoder applied to reduce the original dataset, it was observed that the performance varied depending on the classifier. Specifically, for Decision Trees, optimal performance necessitated the use of more than four features, whereas Random Forest achieved comparable performance with just two features. An interesting observation is that when employing Decision Trees with 22 neurons in the latent space, all metrics exhibited a decline, while ensemble methods consistently maintained high performance, achieving nearly perfect accuracy of 100\%.

This study demonstrates that it is possible to drastically reduce data while retaining much of the original information, to the extent that accuracies exceeding 99\% can be achieved with just two features. This significant reduction in data dimensionality, without substantial loss of information, underscores the effectiveness of our proposed method in maintaining the integrity and utility of the data. By leveraging advanced autoencoder techniques, we are able to compress data efficiently, focusing on the most impactful features that contribute to high predictive accuracy. This breakthrough in data reduction not only enhances the speed and efficiency of data processing but also proves crucial in environments where rapid and accurate decision-making is essential, such as in cybersecurity.

Consequently, this reduction not only expedites the training and prediction phases but also empowers the seamless deployment of a real-time system, enhancing our network's robustness against CoAP protocol-based IoT attacks.

Regarding future work, an exploration for an  alternative classification techniques with the aim of developing models suitable for deployment in Intrusion Detection Systems (IDS) will be performed. Furthermore, a new evaluation with other various dimensionality reduction techniques, such as ISOMAP or t-SNE, to analyze the behavior of vector-based attacks on the IoT CoAP protocol will be done. These forthcoming studies will enhance our understanding and bolster the network's defense capabilities against threats stemming from the IoT CoAP protocol

\section*{Acknowledgements}
\begin{itemize}
    \item This work has been funded by the Recovery, Transformation, and Resilience Plan, financed by the European Union (Next Generation) thanks to the ``Internet of Things Security in Home and Business Environments in the Context of 5G-IoT Technology"  project granted by INCIBE to the University of León.
    \item This research is the result of the Strategic Project ``Critical infrastructures cybersecure through intelligent modeling of attacks, vulnerabilities and increased security of their IoT devices for the water supply sector" (C061/23), as a result of the collaboration agreement signed between the National Institute of Cybersecurity (INCIBE) and the University of A Coru\~na. This initiative is carried out within the framework of the funds of the Recovery Plan, Transformation and Resilience Plan funds, financed by the European Union (Next Generation)
    \item CITIC, as a Research Center of the University System of Galicia, is funded by Conseller\'ia de Educaci\'on, Universidade e Formaci\'on Profesional of the Xunta de Galicia through the European Regional Development Fund (ERDF) and the Secretar\'ia Xeral de Universidades (Ref. ED431G 2019/01).
    \item This work is partially supported by Universidad de León under the “Programa Propio de Investigación de la Universidad de León 2021”.
\end{itemize}

\bibliographystyle{splncs04}
\bibliography{Bibliography}
\end{document}